\documentclass[prd,superscriptaddress,unsortedaddress,twocolumn,showpacs,preprintnumbers,amsmath,amssymb]{revtex4-1}

\usepackage[dvips]{graphicx}
\usepackage{amsmath,amssymb,times}
\usepackage{ulem}

\def\fun#1#2{\lower3.6pt\vbox{\baselineskip0pt\lineskip.9pt
\ialign{$\mathsurround=0pt#1\hfil##\hfil$\crcr#2\crcr\sim\crcr}}}

\newcommand{\beq}{\begin{equation}}
\newcommand{\eeq}{\end{equation}}
\newcommand{\bea}{\begin{eqnarray}}
		  \newcommand{\eea}{\end{eqnarray}}

\DeclareSymbolFont{boldletters}{OML}{cmm} {b}{it}
\DeclareSymbolFontAlphabet{\mathbit}{boldletters}
\DeclareMathSymbol{\alpha}{\mathalpha}{letters}{"0B}
\DeclareMathSymbol{\beta}{\mathalpha}{letters}{"0C}
\DeclareMathSymbol{\gamma}{\mathalpha}{letters}{"0D}
\DeclareMathSymbol{\delta}{\mathalpha}{letters}{"0E}
\DeclareMathSymbol{\epsilon}{\mathalpha}{letters}{"0F}
\DeclareMathSymbol{\zeta}{\mathalpha}{letters}{"10}
\DeclareMathSymbol{\eta}{\mathalpha}{letters}{"11}
\DeclareMathSymbol{\theta}{\mathalpha}{letters}{"12}
\DeclareMathSymbol{\iota}{\mathalpha}{letters}{"13}
\DeclareMathSymbol{\kappa}{\mathalpha}{letters}{"14}
\DeclareMathSymbol{\lambda}{\mathalpha}{letters}{"15}
\DeclareMathSymbol{\mu}{\mathalpha}{letters}{"16}
\DeclareMathSymbol{\nu}{\mathalpha}{letters}{"17}
\DeclareMathSymbol{\xi}{\mathalpha}{letters}{"18}
\DeclareMathSymbol{\pi}{\mathalpha}{letters}{"19}
\DeclareMathSymbol{\rho}{\mathalpha}{letters}{"1A}
\DeclareMathSymbol{\sigma}{\mathalpha}{letters}{"1B}
\DeclareMathSymbol{\tau}{\mathalpha}{letters}{"1C}
\DeclareMathSymbol{\upsilon}{\mathalpha}{letters}{"1D}
\DeclareMathSymbol{\phi}{\mathalpha}{letters}{"1E}
\DeclareMathSymbol{\chi}{\mathalpha}{letters}{"1F}
\DeclareMathSymbol{\psi}{\mathalpha}{letters}{"20}
\DeclareMathSymbol{\omega}{\mathalpha}{letters}{"21}
\DeclareMathSymbol{\varepsilon}{\mathalpha}{letters}{"22}
\DeclareMathSymbol{\vartheta}{\mathalpha}{letters}{"23}
\DeclareMathSymbol{\varpi}{\mathalpha}{letters}{"24}
\DeclareMathSymbol{\varrho}{\mathalpha}{letters}{"25}
\DeclareMathSymbol{\varsigma}{\mathalpha}{letters}{"26}
\DeclareMathSymbol{\varphi}{\mathalpha}{letters}{"27}
\DeclareMathSymbol{\Gamma}{\mathalpha}{letters}{"00}
\DeclareMathSymbol{\Delta}{\mathalpha}{letters}{"01}
\DeclareMathSymbol{\Theta}{\mathalpha}{letters}{"02}
\DeclareMathSymbol{\Lambda}{\mathalpha}{letters}{"03}
\DeclareMathSymbol{\Xi}{\mathalpha}{letters}{"04}
\DeclareMathSymbol{\Pi}{\mathalpha}{letters}{"05}
\DeclareMathSymbol{\Sigma}{\mathalpha}{letters}{"06}
\DeclareMathSymbol{\Upsilon}{\mathalpha}{letters}{"07}
\DeclareMathSymbol{\Phi}{\mathalpha}{letters}{"08}
\DeclareMathSymbol{\Psi}{\mathalpha}{letters}{"09}
\DeclareMathSymbol{\Omega}{\mathalpha}{letters}{"0A}

\begin{document}
%\preprint{SAGA-HE-260-10}
\title{Entanglement between chiral and deconfinement transitions under strong uniform magnetic background field}

\author{Kouji Kashiwa}
\email[]{kashiwa@ribf.riken.jp}
\affiliation{RIKEN Nishina Center, 2-1, Hirosawa, Wako, Saitama 351-0198, Japan}
\date{\today}

\begin{abstract}
Effects of a uniform (electro-) magnetic background field are investigated in the two-flavor nonlocal Polyakov-loop extended Nambu--Jona-Lasinio (PNJL) model.
Temperature-dependences of the chiral order-parameter and the Polyakov-loop are investigated under the magnetic field.
In the nonlocal PNJL model with four-dimensional momentum dependent distribution function, pseudocritical temperatures of the chiral and deconfinement transitions are coincident to each other even if the magnetic field is strong.
This means that we can check the reliability of the distribution function under the strong magnetic field by comparing with lattice QCD data, particularly from the viewpoint of the entanglement between the chiral and deconfinement transitions.
\end{abstract}

\maketitle

%%%%%%%%%%%%%%%%%%%%%%%%%%%%%%%%%%%%%%%%%%%%%%%%%%%%%%%%%%%%%%%%%%%%%%%%%%%
%%%%%  Introduction 
%%%%%%%%%%%%%%%%%%%%%%%%%%%%%%%%%%%%%%%%%%%%%%%%%%%%%%%%%%%%%%%%%%%%%%%%%%%

It is expected that
quantum chromodynamics (QCD) has fruitful phase structures at finite temperature ($T$) and finite real chemical potential ($\mu_\mathrm{R}$) from recent theoretical studies.
The lattice QCD (LQCD) simulation is a powerful method to investigate the QCD thermodynamics at finite $T$ with vanishing $\mu_\mathrm{R}$.
At finite $\mu_\mathrm{R}$, the sign problem comes up in LQCD simulation and then it is not feasible in the $\mu_\mathrm{R}/T > 1$ region; for example, see Ref. \cite{Forcrand} and references therein.
Therefore, model approaches are widely used to investigate the phase structure at finite $\mu_\mathrm{R}$.
Model approaches, however, have the large ambiguities in its foundation; for example, see Ref.~\cite{Kashiwa1}.

Because of the above reasons, we can not obtain reliable QCD phase diagram by individually using LQCD simulation and model approaches at least in the present day.
To overcome this difficulty, imaginary chemical potential matching approach was proposed in Ref.~\cite{Kashiwa2}.
With this approach, we compared model results with LQCD data at finite imaginary chemical potential ($\mu_\mathrm{I}$).
After that, we can obtain reliable extended model and parameters and then we can clarify which interactions are relevant or irrelevant.

In this study, we pay attention to a uniform (electro-) magnetic background field as a situation to check the reliability of the extended model obtained from the imaginary chemical potential matching approach.
Moreover, huge magnetic fields are expected to be generated in heavy ion collisions~\cite{Kharzeev,Skokov}; then, study of the strong magnetic field attracts much more attention.
Recently, effects of the magnetic field were energetically investigated in model calculations.
In Ref.~\cite{Fukushima1}, it is shown that the strength of the magnetic
field has an effect on the chiral crossover but does not on the
deconfinement crossover so much in the Polyakov-loop extended
Nambu--Jona-Lasinio (PNJL) model with smooth regularization.
In Ref.~\cite{Gatto}, the authors consider the entanglement vertex and the eight-quark interaction in the PNJL model with smooth regularization; then the coincidence between the chiral and deconfinement transitions are obtained even if the magnetic field is strong. 
The authors of Ref.~\cite{Gatto} remark that the entanglement vertex is enough to reproduce the coincidence of each transition and it is clearly shown that the entanglement between the transitions is important in the PNJL model
In the linear sigma model coupled to quarks and to Polyakov loops, that coincidence is also investigated~\cite{Mizher}.
These studies suggest that the situation with strong magnetic field provides the important information of the model construction, particularly for the entanglement between the chiral and deconfinement transitions.
In a recent LQCD simulation, it was reported that coincidence of pseudocritical temperatures of the chiral and deconfinement transitions is realized~\cite{D'Elia}.

In this study, we use the PNJL model~\cite{Fukushima2,Ratti,Kashiwa1,Sakai1,Blaschke,Contrera1,Hell1,Contrera2}.
There are different types of the PNJL model and we use the recent PNJL model with the four-dimensional momentum dependent distribution function~\cite{Hell1} in this work.
The nonlocal nature of the PNJL model is natural consequence of QCD nature and thus we do not need the cutoff in the momentum integration in the nonlocal PNJL model in principle.
This is the biggest difference between the local and nonlocal PNJL model and it is an important  advantage if we consider calculations beyond the mean field approximation; for example, see Ref.~\cite{Radzhabov} and references therein. 
Moreover, the chiral and deconfinement transitions are naturally and strongly entangled in the nonlocal PNJL model without any other additional interactions and vertices; for example, see Ref. \cite{Hell1,Hell2, Kashiwa3}.

The Lagrangian density of the two-flavor nonlocal PNJL model is
\begin{align}
{\cal L} &= {\bar q} (i \!\not\!\!D - m_0 ) q + {\cal L}_\mathrm{int}
          - {\cal U} (\Phi[A],{\bar \Phi}[A];T),
\end{align}
where $q$ is the two-flavor quark field, $m_0$ denotes the current quark mass and $D^\nu=\partial^\nu + iA^\nu=\partial^\nu
+i\delta^{\nu}_{0}gA^0_a{\lambda_a / 2}$ with the gauge coupling $g$ and the Gell-Mann matrices $\lambda_a$. 
The last term ${\cal U}$ is called the Polyakov-loop effective potential.
In this study, we treat $A^0$ as the mean field value as
$A_4 = iA^0 = T {\rm diag}(\phi_a,\phi_b,\phi_c)$.
The interaction part is expressed as  
\begin{align}
{\cal L}_\mathrm{int} (x) 
&= G_\mathrm{s} j_a (x) j_a (x), \\
j_a (x) 
&= \int d^4z~ {\cal C}(z) {\bar q} \Bigl( x + \frac{z}{2} \Bigr) \Gamma_a 
                  q \Bigl( x-\frac{z}{2} \Bigr),
\end{align}
where $\Gamma_a$ is 
$\Gamma_{a=(0,\dots,3)}=(1,i\gamma_5 {\vec \tau})$ for the scalar and the pseudoscalar interaction, respectively.
The function ${\cal C}(z)$ is the distribution function here.  
In this study, we use the following distribution function proposed in Ref.~\cite{Hell1} as
\begin{align}
{\cal C} (p^2) = {\cal C} (\omega_n^2,{\bf p}^2) = \left\{
\begin{array}{c}
~~~e^{- d_C^2 p^2} ~~~ (p^2< \Gamma^2)\\
{\cal N} \frac{\alpha_s(p^2)}{p^2} ~~~~(p^2 \geq \Gamma^2) \\
\end{array}
\right.
\label{eq:form_factor}
\end{align}
where $p^2 = (\omega_n)^2 + {\bf p}^2$ at finite $T$ with fermion Matsubara frequency $\omega_n = (2n + 1)\pi T$. 
The function $\alpha_s (p^2)$ means the running coupling constant. 
Constants $d_C$ and ${\cal N}$ are automatically determined to smoothly connect the function at $p^2 = \Gamma^2$ if we set the actual value of $\Gamma$.
It is shown that if we accept the bare $\mu$-dependence of the distribution function by using following replacement of $p_0$,
\begin{align}
p_0 \to i \omega_{n} + \mu + i A_4,
\label{p0}
\end{align}
the $A_4$-dependence of ${\cal C}$ is a natural consequence from existence of the Roberge-Weise periodicity~\cite{RW} as shown in Ref.~\cite{Kashiwa3}.
The $A_4$-dependence is also shown from QCD based analysis~\cite{Kondo}.
In this study, we neglect the $Z$ factor that comes from the quark wave function renormalization to make our discussion simple.
Details of the $Z$ factor in the present nonlocal PNJL model are shown in 
Ref.~\cite{Hell2,Kashiwa3}.

In this study, we use the following logarithmic Polyakov-loop effective potential ~\cite{Hell1} as
\begin{align}
{\cal U}
&= T^4 \Bigl[ -\frac{1}{2} b_2(T) {\bar \Phi} \Phi
\nonumber\\
&             + b_4(T) \ln\{ 1 - 6 {\bar \Phi} \Phi 
                            +4 ({\bar \Phi}^3 + \Phi^3) 
                             -3 ({\bar \Phi}\Phi)^2 \} \Bigr], \\
b_2 &= a_0 + a_1 \Bigl( \frac{T_0}{T} \Bigr)
           + a_2 \Bigl( \frac{T_0}{T} \Bigr)^2
           + a_3 \Bigl( \frac{T_0}{T} \Bigr)^3,
\\
b_4 &= a_4 \Bigl( \frac{T_0}{T} \Bigr)^3,
\end{align}
where the $\Phi$ and ${\bar \Phi}$ with the Polyakov-gauge are
$\Phi =
\Bigl( e^{i \phi_a} + e^{i \phi_b} + e^{i \phi_c} \Bigr)/3,
~
{\bar \Phi} 
 = \Phi^* $
here
$\phi_a = ( \phi_3 + \phi_8/\sqrt{3} )/T$,
$\phi_b = ( -\phi_3 + \phi_8/\sqrt{3} )/T$ and 
$\phi_c = - (\phi_a + \phi_b)$.

From here, we consider the uniform magnetic background field 
$A_\mu^{ext}$. 
The magnetic field along the positive $z$ axis is expressed as
$A_\mu^{ext} = B x_1 \delta_{\mu 2}$ where we choice the Landau gauge.
In this case, the four-dimensional momentum becomes
$
p  = (p_0,0,{\cal Q}\sqrt{2k |Q_f e B| },p_z)$
and then 
$
{\bar p}^2 = p_z^2 + 2 k |Q_f eB|
$
below where $k$ labels each Landau level and ${\cal Q}$ denotes the sign function as ${\rm sgn}(Q_f)$ with electric-charge $Q_f$ for each quark flavor $f=u,d$.
Making the mean field approximation and introducing Landau levels, the thermodynamical potential can be expressed as
\begin{align}
\Omega 
&= G_s \sigma^2 + {\cal U} 
\nonumber\\
&-
T \frac{|Q_f e B|}{2\pi}
\sum_{f}
\sum_{i}
\sum_{n=-\infty}^\infty 
\sum_{k=0}^\infty \beta_k 
\int_{-\infty}^\infty \frac{dp_z}{2\pi}
\nonumber \\
&\times \ln \Bigl[ \beta^2 
    \Bigl\{ \Bigl( \omega_{n,i} \Bigr)^2
          + \Bigl( E_f^i({\bar p}^2) \Bigr)^2 \Bigr\} \Bigr],
\label{TP-MF}
\end{align}
where 
$\beta=1/T$,
$E_f^{i} = \sqrt{{\bar p}^2 + (M_f^i)^2}$,
$\omega_{n,i} = (2 n + 1) \pi T + (\phi_i T - i \mu)$
and $M={\rm diag} (M_u,M_d)$ with
$M_u^i=M_d^i = m_0 - 2G_s {\cal C}_i({\bar p}^2) \sigma$ 
here $i$ represents the color indices.
It is shown that the $\mu$ is set to zero in this calculation.
The coefficient $\beta_k = 2 - \delta_{k0}$ express the degeneracy of the Landau level. 
The thermodynamical potential (\ref{TP-MF}) is difficult to converge with numerical summation over $n$.
Therefore, we rewrite it as
\begin{align}
\Omega
&= 
G_s \sigma^2 + {\cal U} 
+ \Omega_{\rm 0}
\nonumber\\
&- 
T \frac{|Q_f e B|}{2\pi}
\sum_{f}
\sum_{i}
\sum_{n} 
\sum_{k} \beta_k 
\int_{-\infty}^\infty \frac{dp_z}{2\pi}
\nonumber\\
&\times
\ln \Bigl[ 
\frac{
       \omega_{n,i}^2
     + \{ E_f^i({\bar p}^2) \}^2 }
     {
       \omega_{n,i}^2
     + {\bar p}^2 + m_0^2 }
\Bigr],
\end{align}
where $\Omega_{\rm 0}$ means the thermodynamical potential with noninteracting quarks but it is affected by the uniform background field $A_4$ and $A^{ext}$. 
The final form is
\begin{align}
\Omega_{\rm 0} 
&=
- 
2T \frac{|Q_f e B|}{2\pi}
\sum_{n} 
\sum_{k} \beta_k 
\int^\infty_{-\infty} \frac{dp_z}{2\pi}
\Bigl[ \ln f + \ln {\bar f} \Bigr],
\label{Omega0}
\\
f &= 
1+3(\Phi + {\bar \Phi}e^{- \beta e^-({\bar p})})
                      e^{- \beta e^-({\bar p})} 
         + e^{- 3\beta e^-({\bar p})},
\\
{\bar f} &= 
1+3({\bar \Phi} + \Phi e^{- \beta e^+({\bar p})})
                       e^{- \beta e^+({\bar p})} 
         + e^{- 3\beta e^+({\bar p})},
\end{align}
with $e^\pm = \sqrt{{\bar p}^2+m_0^2} \pm \mu$.
The vacuum part of $\Omega_0$ is already subtracted from thermodynamical potential (\ref{Omega0}) because it does not affect the position of the global minimum.

The present nonlocal PNJL model has three parameters in the NJL-part, $G_s$, $m_0$ and $\Gamma$.
The Polyakov-loop effective potential has six parameters, $a_0$, $a_1$, $a_2$, $a_3$, $a_4$ and $T_0$.
Parameters in the Polyakov-loop effective potential are fixed by using LQCD data in the pure-gauge limit.
The actual values are $a_0=3.51$, $a_1=-2.56$, $a_2=15.2$, $a_3=-0.62$ and $a_4=-1.68$.
The energy scale $T_0$ is set to $270$ MeV.
Parameters in the NJL-part are fixed by using empirical values of the pion mass and its decay constant. 
Actual values are $m_0=3.3$ MeV, $G_\mathrm{s} = 20.6$ GeV$^{-2}$ and $\Gamma = 0.827$ GeV.

First, we show results of the PNJL model with smooth regularization to compare with the nonlocal PNJL model with a four-dimensional distribution function.
Here, we do not present actual formalism of it; see Ref.~\cite{Gatto} for the details of the formalism and the parameter set.
Below, we call the PNJL model with smooth regularization and the four-dimensional distribution function as 3D-PNJL model and 4D-PNJL model, respectively.
%%%%%%%%%%%%%%%% Fig %%%%%%%%%%%%%%%%%%%%%
\begin{figure}[htbp]%[H]
\begin{center}
 \includegraphics[width=0.231\textwidth]{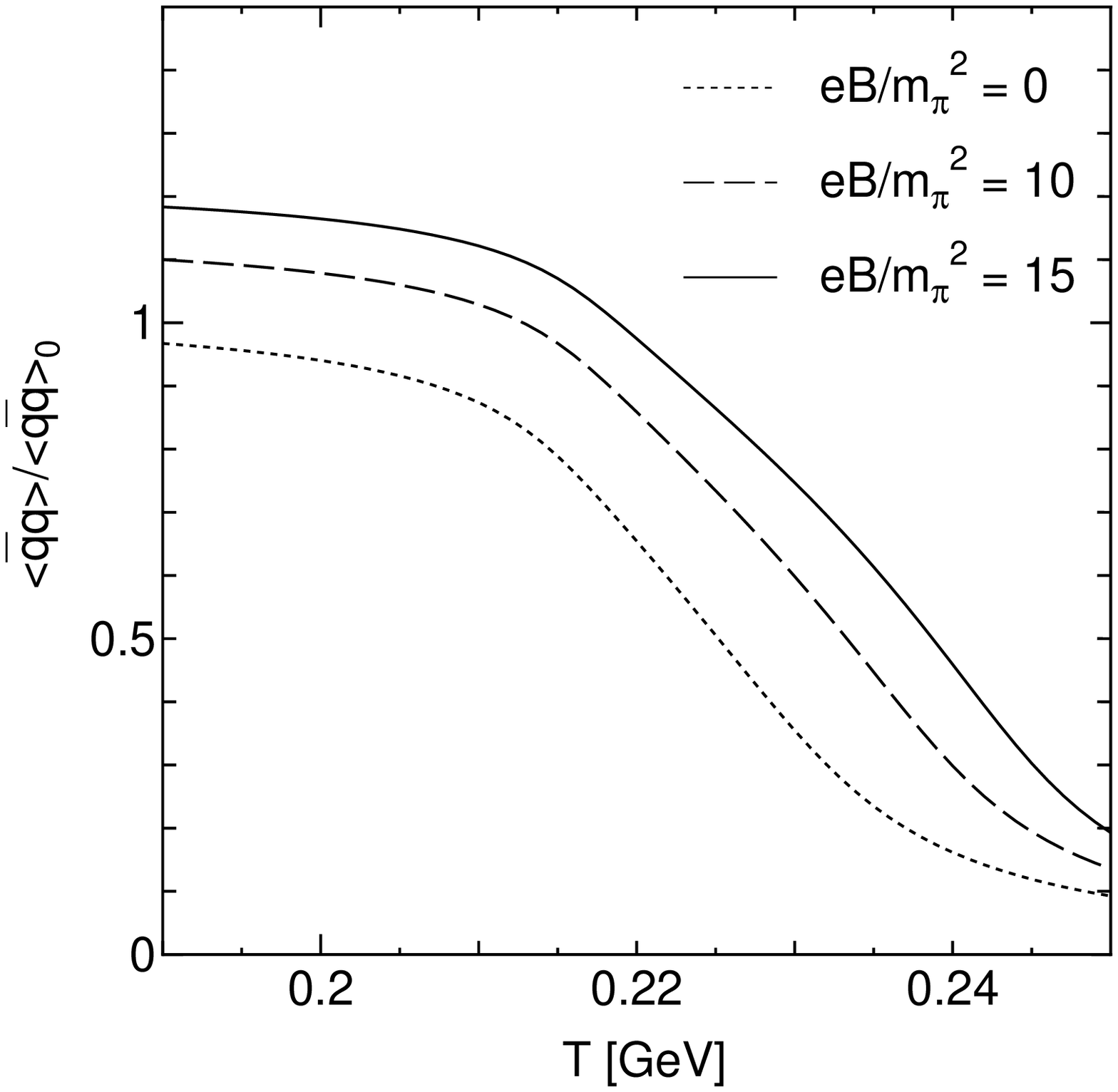}
 \includegraphics[width=0.235\textwidth]{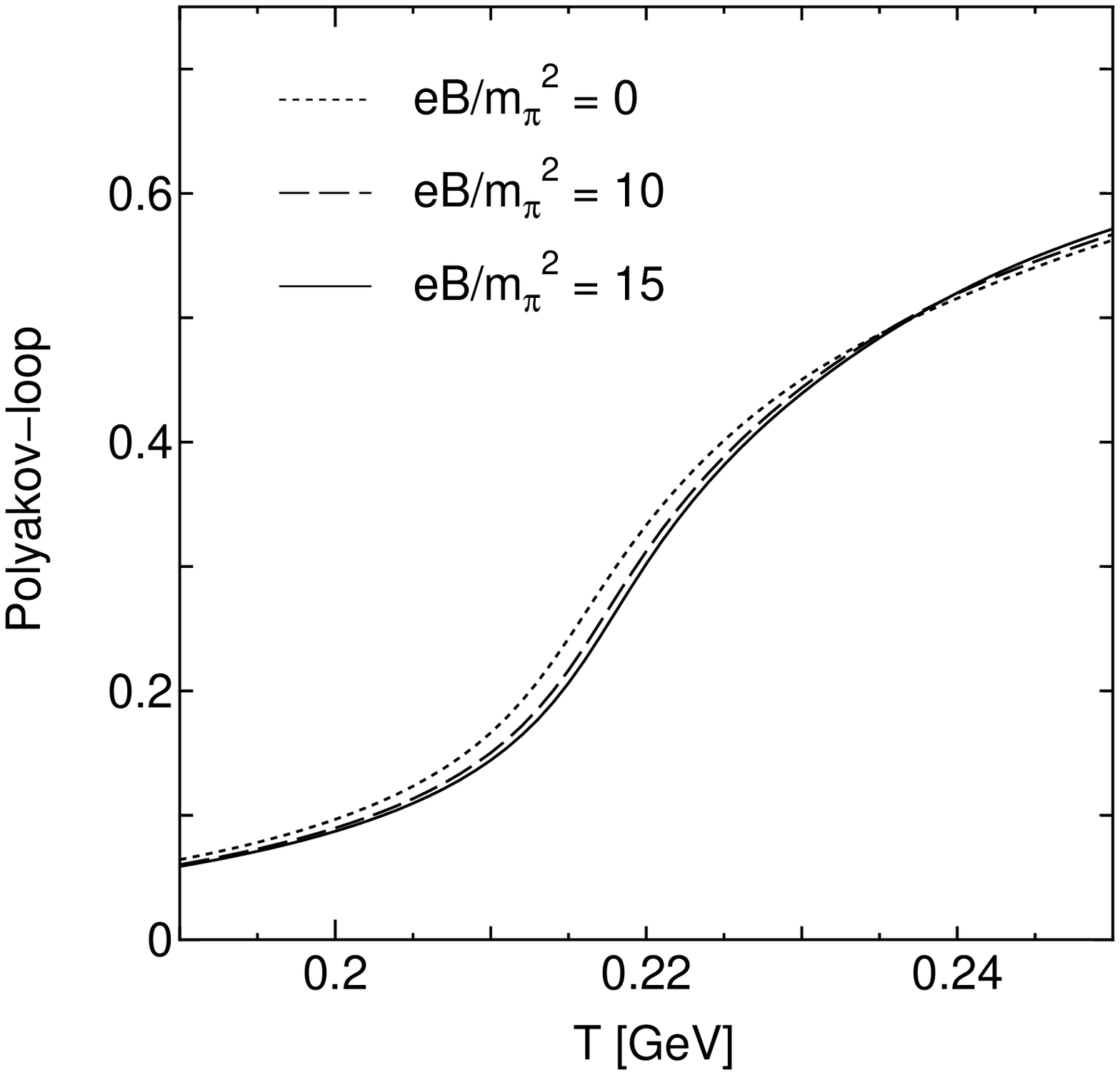}
\end{center}
\caption{ The $T$ dependence of the chiral condensate and the Polyakov-loop in the 3D-PNJL model for $eB/m_\pi^2=0$, $10$ and $15$, respectively.
The chiral condensate is normalized by that at $T=0$ with $eB=0$.
}
\label{Fig:LPNJL-a}
\end{figure}
%%%%%%%%%%%%%%%%%%%%%%%%%%%%%%%%%%%%%%%%%%  
Figure~\ref{Fig:LPNJL-a} shows the $T$ dependence of the chiral condensate and the Polyakov loop in the 3D-PNJL model for $eB/m_\pi^2=0$, $10$ and $15$.
In the case of the 3D-PNJL model, the Polyakov loop is less sensitive than the chiral condensate against the strength of the magnetic field.
This means that the difference between pseudocritical temperatures becomes large when the magnetic field becomes strong.
Rough estimation of pseudocritical temperatures are summarized in Table~\ref{Table-PCT}.
In the estimation, we use the temperature-derivative of order parameters,
and the numerical error is about $\pm0.001$ GeV.

%%%%%%%%%%%%%%%%%%%%%%%%%%%%%%%%%%%%%%%%%%%%%%%%%%%%%%%%%%%%%%%%%%%%%%%%%%%
% Pseudo-critical temperatures %
%%%%%%%%%%%%%%%%%%%%%%%%%%%%%%%%
\begin{table}[h]
\begin{center}
\begin{tabular}{cccc}
\hline
$eB/m_\pi^2$       & ~~~~$0$~~~~  & ~~~~$10$~~~~ & ~~~~$15$~~~~  \\
\hline
\hline
$T_\chi$ (3d-PNJL) & 0.226  & 0.236 & 0.241 \\
\hline
$T_d$ (3d-PNJL)    & 0.216  & 0.217 & 0.218 \\
\hline
\hline
$T_\chi$ (4d-PNJL) & 0.208  & 0.215 & 0.221  \\
\hline
$T_d$ (4d-PNJL)    & 0.208  & 0.215 & 0.221 \\
\hline
\end{tabular}
\caption{Rough estimation of pseudo-critical temperatures for the chiral
 ($T_\chi$) and the deconfinement transition ($T_d$) determined from the peak position of the temperature derivative of order parameters.
Unit of $T_\chi$ and $T_d$ is GeV and these have about $\pm0.001$ GeV numerical error.}
\label{Table-PCT}
\end{center}
\end{table}
%%%%%%%%%%%%%%%%%%%%%%%%%%%%%%%%%%%%%%%%%%%%%%%%%%%%%%%%%%%%%%%%%%%%%%%%%%%

%%%%%%%%%%%%%%%% Fig %%%%%%%%%%%%%%%%%%%%%
\begin{figure}[htbp]%[H]
\begin{center}
 \includegraphics[width=0.230\textwidth]{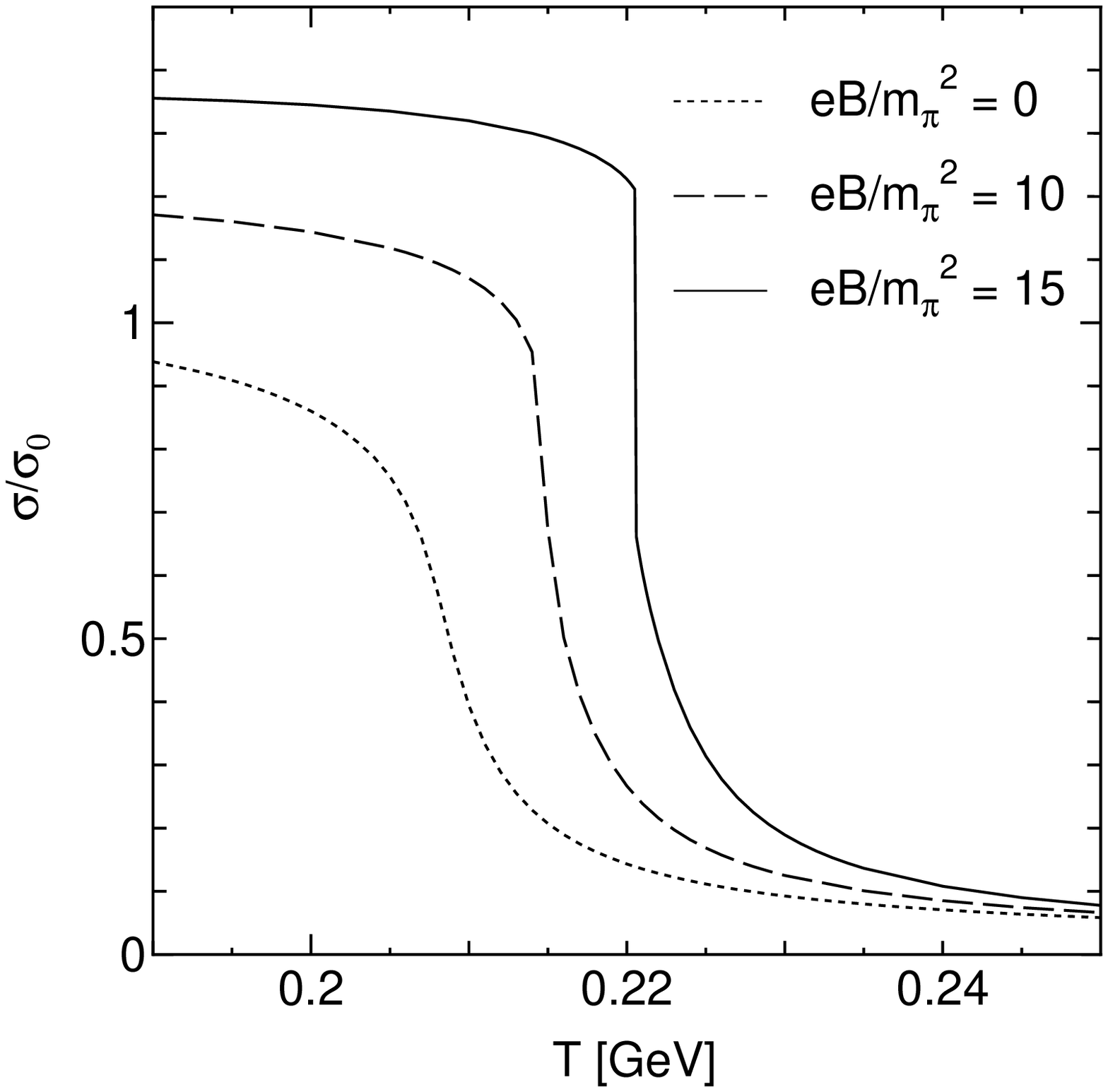}
 \includegraphics[width=0.235\textwidth]{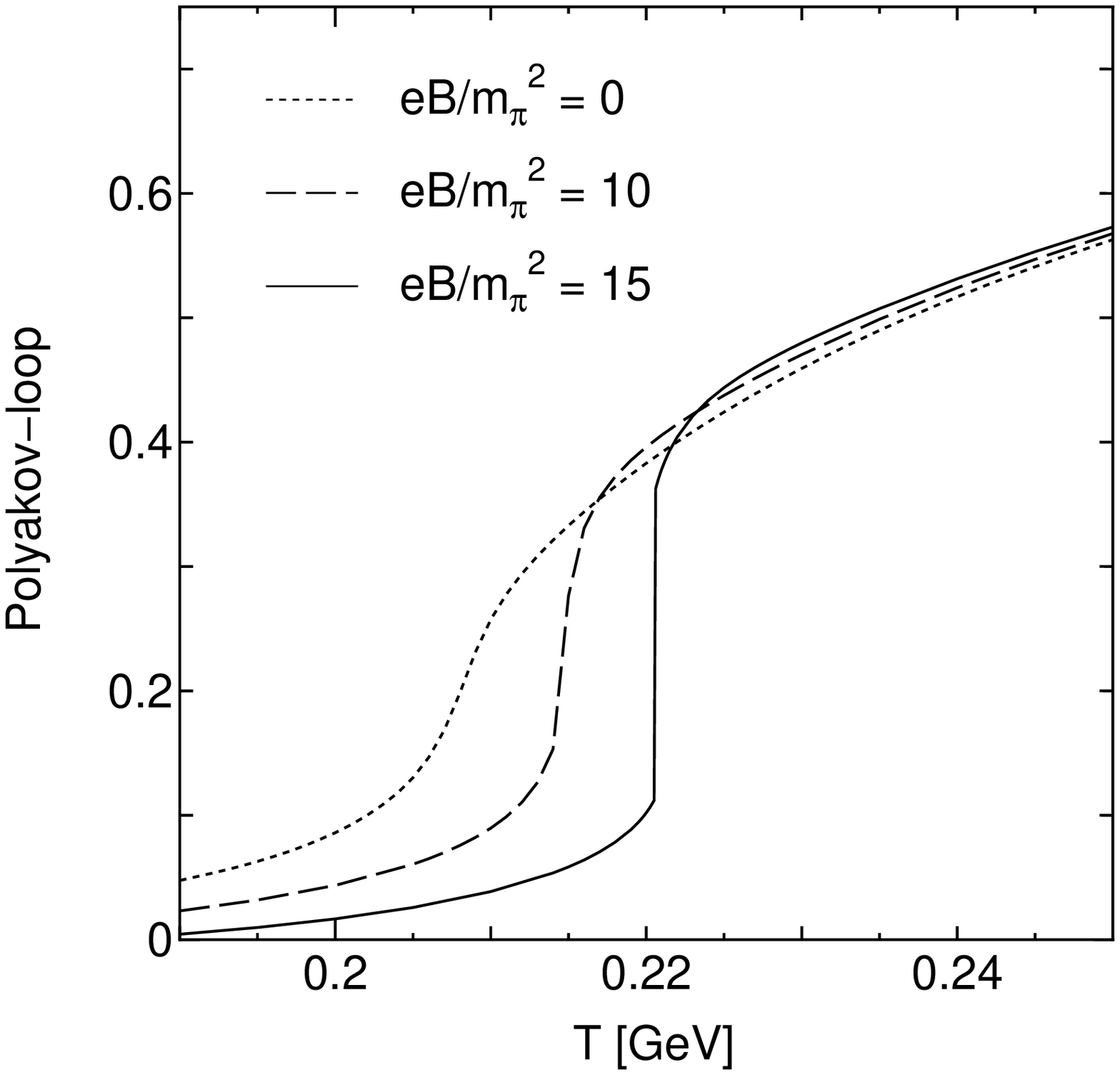}
\end{center}
\caption{ The $T$ dependence of the chiral order parameter ($\sigma$) and the Polyakov-loop in the 4d-PNJL model for $eB/m_\pi^2=0$, $10$ and $15$, respectively.
The chiral order-parameter is normalized by that at $T=0$ with $eB=0$.
}
\label{Fig:NPNJL-a}
\end{figure}
%%%%%%%%%%%%%%%%%%%%%%%%%%%%%%%%%%%%%%%%%%  
Next, we show results of the 4D-PNJL model.
Figure~\ref{Fig:NPNJL-a} show the $\sigma$ field and the Polyakov loop for $eB/m_\pi^2=0$, $10$ and $15$.
The $\sigma$ field can be considered the chiral order parameter which is the indicator of the chiral transition with finite quark current mass because it appears in the quark mass function.
In the case of the 4D-PNJL model, the pseudocritical temperatures of the chiral and deconfinement transitions are very close to each other at several $eB/m_\pi^2$ -- the same as te LQCD suggestion and actual values are also shown in Table~\ref{Table-PCT}.
Moreover, we can see that the transitions are close to the first-order behavior when the magnetic field becomes strong.
This first-order tendency is already suggested in LQCD simulation
~\cite{D'Elia}.
The 4D-PNJL model can naturally reproduce the LQCD prediction without some other additional interactions.
At $\mu=0$ with $eB=0$, transitions are considered as the crossover or the very weak first-order by recent LQCD data; then it is difficult to clearly check the reliability of the entanglement effect in the model.
However, the transitions become sharp first order under a strong magnetic field and then it is possible to clearly check the reliability of the model through the dependence of the model against the strength of the uniform magnetic background field. 

\begin{figure}[htbp]%[H]
\begin{center}
 \includegraphics[width=0.28\textwidth]{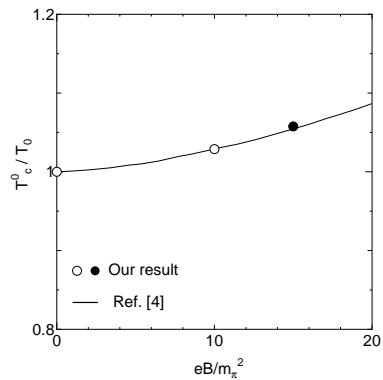}
\end{center}
\caption{
The percentage of increase of the pseudocritical and critical temperatures as a function of $eB/m_\pi^2$.
Here, $T_c$ means the pseudocritical or critical temperatures for each $eB/m_\pi^2$
and $T^0_c$ means the pseudocritical temperature with $eB/m_\pi^2=0$.
The solid line represents the EPNJL model results shown in Fig.~6 of Ref.~\cite{Gatto}.
Opened and closed symbols are the pseudocritical and critical temperatures of 4D-PNJL model, respectively.}
\label{Fig:ratio}
\end{figure}

Finally, we discuss our results with the data in Ref.~\cite{Gatto}.
Figure~\ref{Fig:ratio} shows the percentage of increase of the pseudocritical and critical temperatures as a function of $eB/m_\pi^2$.
In the figure, $T_c$ means the pseudocritical or critical temperatures for each $eB/m_\pi^2$ where $T^0_c$ shows the pseudocritical temperature with $eB/m_\pi^2=0$.
The solid line represents the results of the EPNJL model picked up from Fig.~6 in Ref.~\cite{Gatto} and this model corresponds to the 3D-PNJL model with the entanglement vertex.
We only plot the pseudocritical temperature of the chiral transition because pseudo-critical temperatures of the chiral and deconfinement transitions are very close to each other in the EPNJL model.
Opened and closed symbols are the pseudocritical and critical temperatures of the 4D-PNJL model, respectively.
The 4D-PNJL and 3D-PNJL models show good agreement and the crossover is changed into the first-order above $eB/m_\pi^2 \sim 11$ in the present 4D-PNJL model.
%%%%%%%%%%%%%%%% Fig %%%%%%%%%%%%%%%%%%%%%
%%%%%%%%%%%%%%%%%%%%%%%%%%%%%%%%%%%%%%%%%%  

In the 4D-PNJL model, the form (\ref{p0}) is the natural (minimal) form by respecting the replacement manner of $p_0$, the preservation of properties of QCD at finite imaginary chemical potential~\cite{RW} and manifestation of the recent QCD based analysis~\cite{Kondo} when we accept the bare $\mu$ dependence in the distribution function.
Importance of this form is also expressed in the functional renormalization group analysis~\cite{Braun}.
Therefore, there is the possibility that we can indirectly investigate the
$\mu$ dependence of the model even at $\mu=0$ through Eq.~(\ref{p0}).
Of course, the form (\ref{p0}) is a sufficient but not necessary condition.
Therefore, we must carefully investigate the $\mu$ dependence step by step.

We can use the system with the strong uniform magnetic background field as the situation to quantitatively check the model reliability in the near future when more accurate LQCD data are obtained.
\\

K. K. thanks W. Weise and T. Hell for useful discussion about the nonlocal PNJL model.
K. K. also thanks K. Fukushima and M. Ruggieri for their comments and
H. Kouno and M. Yahiro for their careful reading of this manuscript. 
This work is supported by RIKEN and partially supported by the Japan Society for the Promotion of Science for Young Scientists.

%%%%%%%%%%%%%%%%%%%%%%%%%%%%%%%%%%%%%%%%%%%%%%%%%%%%%%%%%%%%%%%%%%%%%%%%%%%%%%%%

\end{document}